# Promisedland: An XR Narrative Attraction Integrating Diorama-to-Virtual Workflow and Elemental Storytelling

Xianghan Wang*
*Integrated Digital Media*
New York University
New York, the United States
xw2264@nyu.edu
*Corresponding author

Chingshuan Hsiao
*Integrated Digital Media*
New York University
New York, the United States
csh444@nyu.edu

Shimei Qiu*
*Human Computer Interaction*
Rochester Institute of Technology
Rochester, United States
sq3465@rit.edu
*Corresponding author

*Abstract*— Promisedland is a mixed-reality (MR) narrative attraction that combines cultural storytelling, ecological education, and an innovative hybrid production workflow. Set in a future Earth suffering from elemental imbalance, users embark on an interactive journey guided by symbolic characters to restore harmony through the collection of five classical elements: metal, wood, water, fire, and earth. To prototype this experience, we introduce a low-cost, high-fidelity Diorama-to-Virtual pipeline—handcrafting physical scale models, 3D scanning, and integrating them into Unreal Engine. This process enables rapid spatial prototyping while preserving the material expressiveness and narrative consistency of the physical environment. To further enhance immersion, the experience incorporates a Stewart Platform to provide motion feedback synchronized with the virtual ride dynamics, reinforcing spatial presence and embodied engagement. The final prototype runs on Meta Quest, supporting dynamic interactions and real-time visual feedback. Promisedland offers a replicable design blueprint for future XR narrative installations across museums, cultural exhibitions, and themed entertainment. It proposes a new framework for XR Narrative Attractions—where physical and digital elements converge to deepen immersion, agency, and emotional engagement.

## I. Introduction

### A. Industry Gaps

Traditional theme park attractions have always relied on expensive, meticulously crafted physical sets, designed primarily for visual spectacle rather than interactive engagement. In these conventional setups, visitors are typically positioned as passive observers who experience rides and installations from a fixed viewpoint with limited opportunities for agency or meaningful interaction. Such experiences, while visually impressive, often fail to leverage visitors' active participation or to create emotionally resonant, memorable narratives. A great majority of the most visited theme parks in the sample have thematic and narrative characteristics, which provide appeal to visitors [1]. However, these are often not widely accessible, require long development cycles, and tend to limit interaction to short time windows. Moreover, such interactions are frequently designed for excitement or thrill-seeking rather than for educational enrichment or narrative depth, further underscoring the need for more inclusive, flexible, and purpose-driven engagement models in themed entertainment.

In recent years, the adoption of Virtual Reality (VR), Augmented Reality (AR), and Mixed Reality (MR)—collectively referred to as Extended Reality (XR)—has introduced new possibilities for immersive storytelling and user interaction within themed attractions and exhibition spaces. However, many existing XR-based installations continue to emphasize technological novelty and visual spectacle at the expense of narrative depth and cultural relevance [2]. Consequently, despite their interactive potential, these XR experiences frequently lack a coherent narrative structure, resulting in superficial user engagement and diminished emotional impact.

Moreover, many current XR installations, despite incorporating interactive components, still exhibit a noticeable gap between user interactions and meaningful narrative outcomes [3]. Interactive choices often appear inconsequential to the overall storyline, leading users to perceive their involvement as insignificant or arbitrary. This disconnect undermines the potential of XR to create immersive, story-driven experiences that can engage users on deeper emotional and intellectual levels.

Existing XR installations in entertainment and cultural spaces often rely on prefabricated digital assets and generic interactivity, frequently prioritizing technological features and visual spectacle over narrative coherence and intuitive user interfaces. This approach has been shown to result in decreased visitor engagement and diminished immersion [4]. For instance, a recent industry survey revealed that 52% of theme park visitors reported technical difficulties when engaging with screen-based interactive experiences, contributing further to visitor frustration and reduced immersion [5]. Additionally, comparative research on VR exhibitions indicates that reality-based spatial environments foster greater user communication and emotional connection compared to abstract or heavily virtualized ones [6]. Our Promisedland prototype addresses these gaps specifically by integrating physical memory triggers and narrative-driven interactions, combining tactile familiarity with immersive storytelling to enhance emotional and educational outcomes.

To address these shortcomings, there is a clear need for new frameworks and methodologies that effectively integrate compelling narratives, meaningful interaction, and innovative production methods. Such frameworks should emphasize not only technical achievement but also the cultural resonance and educational value inherent in spatial storytelling.

*B. Key Contributions*

This paper introduces Promisedland, a novel mixed-reality (MR) narrative attraction designed to bridge the identified industry gaps by emphasizing narrative coherence, cultural resonance, and meaningful interaction within XR installations. Specifically, the contributions of this research are threefold:

First, we propose and exemplify a new design concept—the "XR Narrative Attraction"—which tightly integrates interactive gameplay and immersive storytelling through culturally grounded frameworks. By leveraging the Eastern philosophical concept of the five elements (metal, wood, water, fire, and earth), this project showcases how abstract cultural narratives can effectively drive user engagement, enrich emotional immersion, and provide meaningful ecological education.

Second, we introduce an innovative hybrid production pipeline—the Diorama-to-Virtual workflow—that combines traditional handcrafted scale models, digital scanning techniques, and virtual environment reconstruction in Unreal Engine 4.27. This pipeline offers a cost-effective, rapid prototyping alternative to conventional digital modeling, enabling small creative teams to maintain narrative consistency, spatial logic, and tactile realism throughout the development process.

Third, the project presents a scalable and modular MR prototype with significant potential for broader applications. By incorporating synchronized motion feedback through a Stewart Platform and deploying the experience on Meta Quest 2, our prototype serves as a versatile template for future XR installations across various contexts, including museums, cultural exhibitions, educational programs, and themed entertainment venues.

Overall, these contributions establish a replicable framework that not only advances narrative-driven XR experiences but also enhances the practical accessibility of immersive storytelling design for diverse creators and institutions.

## II. RELATED WORK

Despite significant advancements in XR technologies, current commercial installations often prioritize technological novelty over meaningful narrative integration. For example, immersive experiences such as those offered by Sandbox VR emphasize visual spectacle and interactive elements but frequently lack coherent storytelling, resulting in superficial engagement and limited emotional resonance [7]. Comparative research on interactive cultural heritage also highlights that meaningful narrative structures and culturally coherent content are essential to fostering deeper emotional connections in immersive experiences [8]. Additionally, recent studies emphasize that immersive storytelling in XR greatly benefits from coherent narrative integration, significantly enhancing emotional engagement and user presence compared to experiences primarily driven by technological novelty or spectacle alone [9]. This gap between technical capability and narrative depth highlights a broader challenge: designing XR experiences that leverage interactivity while effectively embedding culturally grounded narratives.

Hybrid production pipelines—combining physical modeling with virtual reconstruction—offer a promising approach to address this challenge. Such methods have already demonstrated substantial value in fields like film previsualization and immersive exhibition design. For instance, films such as Blade Runner 2049 employed physical miniature sets extensively to capture nuanced details and authentic textures that are difficult to achieve through purely digital CGI, thus significantly enhancing narrative immersion and tactile realism [10]. Yet, within XR content creation, hybrid methods remain underutilized, as the field typically favors purely digital workflows for perceived efficiency advantages. However, research shows that combining handcrafted physical models with digital scanning and modeling can significantly improve spatial prototyping, narrative coherence, and emotional engagement in XR experiences [11]. Additionally, comparative studies indicate that reality-based spatial environments—leveraging tangible, familiar elements—can significantly enhance user communication, emotional engagement, and educational effectiveness compared to heavily abstracted or virtualized spaces [12]. Building on these insights, our research introduces Promisedland, a culturally enriched hybrid XR attraction employing a practical Diorama-to-Virtual pipeline. This approach not only addresses existing methodological gaps but also provides a replicable, narrative-driven framework for diverse XR storytelling applications.

## III. DESIGN GOALS AND NARRATIVE INTERACTION FRAMEWORK

*A. Core Design Objectives*

Promisedland was developed with a strong educational intent, emphasizing the ecological interconnectedness of our natural world. By structuring interactive quests around the classical Eastern philosophy of the five elements—metal, wood, water, fire, and earth—we designed scenarios in which participants' decisions and actions directly shape narrative outcomes. This design approach explicitly highlights ecological principles, encouraging participants to recognize their role and responsibility within environmental systems, thus fostering deeper ecological awareness and engagement [13].

To effectively achieve this educational intent, the experience is set in a near-future Earth devastated by ecological collapse, situating participants as lunar explorers tasked with restoring planetary balance through the retrieval of five critical elemental energies. To reinforce emotional immersion, we introduced two symbolic virtual guides: a fairy named Fay, representing natural guidance and harmony, and a butterfly-shaped vehicle named Vanny, symbolizing transformation, fragility, and renewal. By integrating character-driven interactions, spatial exploration, and meaningful decision-making, the experience seeks to cultivate profound emotional connections, immersing participants deeply in the unfolding ecological narrative.

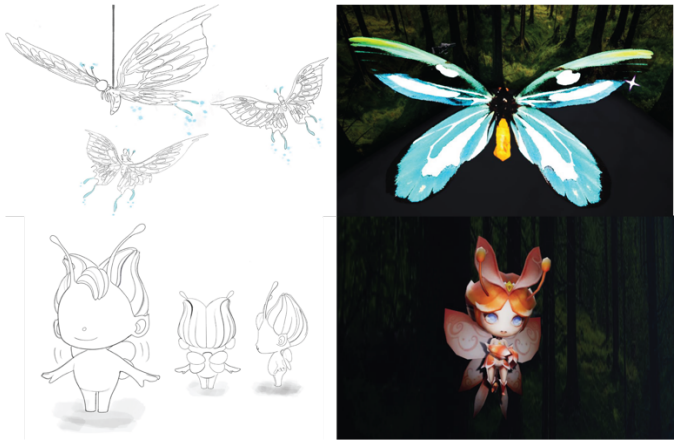

Fig. 1. Fairy Guider and Ride Vehicle

Recognizing the practical constraints of small creative teams, the development of Promisedland also emphasizes accessibility and efficiency by employing an innovative Diorama-to-Virtual workflow. This hybrid production method integrates handcrafted physical models, rapid 3D scanning techniques, and virtual scene reconstruction in Unreal Engine. Such an approach allows efficient prototyping and iterative development without substantial financial or technical resources, significantly lowering entry barriers for XR narrative creation, thereby promoting greater accessibility and replicability for various creators and institutions.

*B. Attraction Layout and User Journey Design*

To optimize the user experience and efficiently utilize available spatial constraints, we adopted a "Spaghetti Layout" for the attraction's physical and virtual design. This layout, characterized by winding pathways and strategic spatial overlaps, enhances curiosity, discovery, and engagement by continuously revealing new environments as participants progress through the journey [14]. Such spatial arrangements not only intensify immersion but also reinforce the exploratory nature inherent in narrative-driven XR experiences.

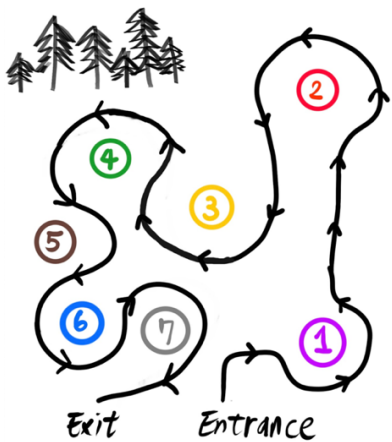

Fig. 2. Spaghetti Ride Layout

Within this carefully structured spatial framework, the Promisedland experience comprises seven key scenes: an onboarding introduction, five elemental-themed challenges, and a concluding scene. Upon entering the experience, participants first encounter an onboarding scenario designed to familiarize them with interactive mechanics and storytelling context. They begin by overcoming a preliminary challenge involving a mutant creature, followed by meeting Fay, who provides voice-guided instructions, narrative context, and emotional engagement. Subsequently, participants sequentially navigate through the elemental realms in the order of fire, earth, metal, wood, and water. Each successfully completed challenge results in the collection of an elemental energy, cumulatively restoring ecological balance and shaping the narrative towards a harmonious or catastrophic outcome.

*C. Elemental Interaction Mechanics and Challenges*

Each elemental environment within Promisedland is carefully designed with distinct thematic interactions and ecological messaging:

- Fire: Participants experience a volcanic eruption scenario, tasked with dynamically dodging streams of lava and environmental hazards while collecting fragments of fire elements. This scenario vividly portrays ecological disruption caused by climate change, emphasizing urgency and awareness.

- Metal: Participants encounter a fierce dragon guardian, representing challenges associated with resource extraction and environmental exploitation. Successfully defeating this guardian underscores the careful balance needed when utilizing finite resources.

- Wood: Users traverse an unstable forest environment, actively avoiding falling logs and debris resulting from severe deforestation. This interactive mechanism starkly illustrates the consequences of unchecked environmental destruction and the critical need for sustainable practices.

- Earth: In a scene characterized by a severe sandstorm enveloping a deserted city, users must navigate through environmental obstacles, symbolizing widespread desertification and soil degradation, highlighting the urgency of ecological restoration efforts.

- Water: In an immersive underwater setting, participants engage in restoration activities that purify water sources and rehabilitate marine ecosystems, emphasizing the vital importance of clean water and marine conservation.

Each scenario's interactive mechanics are complemented by real-time visual feedback, responsive particle effects, and continuous narrative guidance from the virtual companions Fay and Vanny, reinforcing ecological themes and deepening participants' emotional engagement.

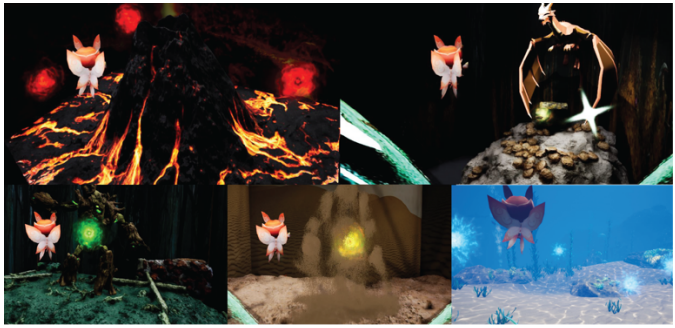

Fig. 3. Five elements scenes in Unreal Engine

## IV. Production Pipeline: Diorama-to-Virtual Workflow

### A. Physical Diorama Construction

To realize the vision of Promisedland, we adopted an innovative Diorama-to-Virtual workflow designed to efficiently translate handcrafted physical environments into immersive virtual scenes, significantly enhancing authenticity and narrative coherence.

In the first stage, we constructed detailed physical dioramas using diverse materials, including black foam board for structural layout and spatial partitions, colored modeling clay to sculpt distinct terrains, artificial moss and grass mats to simulate realistic vegetation, and plastic miniature trees and plants to enrich environmental details. Additional natural materials, such as sand, small stones, wooden sticks, and fabric, were integrated to authentically represent various ecological zones and dynamic environmental effects, including damaged vegetation and disturbed water surfaces. Handcrafting these scenes allowed for material expressivity and set a tangible emotional tone, directly supporting narrative immersion. Furthermore, creating physical dioramas enabled more intuitive spatial planning, facilitating early iteration, quick adjustments, and smoother collaboration within the team.

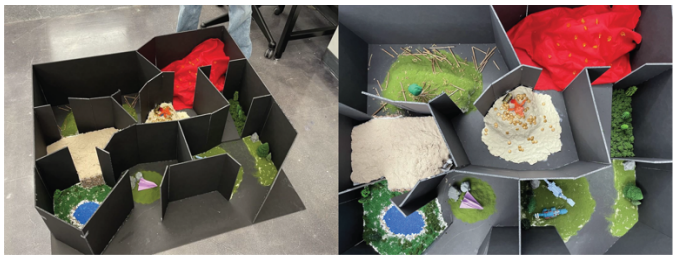

Fig. 4. Physical Diaroma

### B. 3D Scanning & Mesh Optimization

After building the physical solid model, we digitized our physical dioramas through advanced photogrammetry techniques. We utilized Epic Games' RealityScan application to create detailed 3D scans of each individual scene, allowing us to preserve intricate physical details, material textures, and spatial arrangements. Scanning each scene individually enabled the retention of nuanced handmade textures—such as fingerprints, brush strokes, and subtle material imperfections—reinforcing a sense of analog warmth and emotional resonance in the final digital environment. After scanning, we imported the generated 3D models into Blender for mesh optimization, removing excess or redundant meshes, refining geometry, and ensuring efficient real-time rendering performance within Unreal Engine. This careful post-processing guaranteed high fidelity and authenticity, crucial for maintaining narrative coherence and immersive depth in the XR experience.

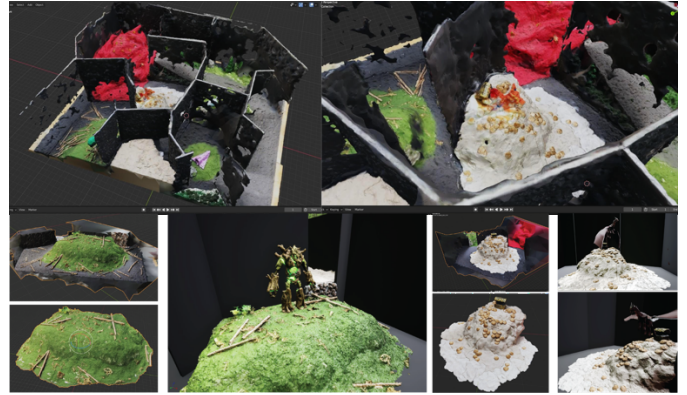

Fig. 5. Examples of 3D scanning and mesh optimization

### C. Integration into Unreal Engine

Following the optimized 3D scanning and mesh preparation processes, we imported the refined digital diorama models into Unreal Engine to construct Promisedland's immersive virtual environment. Each scene was meticulously reconstructed with precise collision meshes, interactive triggers, and comprehensive animation state machines to enable real-time user interaction while ensuring seamless narrative continuity.

We designed the virtual ride experience using Unreal's spline system to create accurate movement trajectories and camera paths that simulate authentic ride dynamics. The Unreal Sequencer tool allowed us to choreograph key animation sequences, guiding users through the elemental scenes with synchronized environmental transformations and interactive prompts. To further enhance visual immersion, we implemented a sophisticated multi-layered visual effects (VFX) system utilizing Unreal's Niagara particle system. Particle effects simulated dynamic elemental phenomena such as volcanic lava flows, water splashes, and energy-driven visual transitions, effectively reinforcing the immersive ecological narrative.

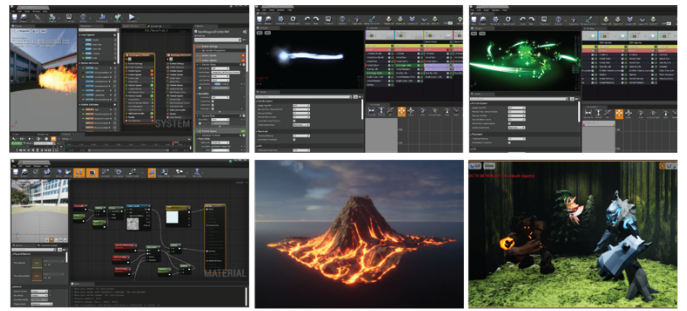

Fig. 6. VFX system in Unreal Engine

The final integration seamlessly unified scanned assets, procedural Niagara-based VFX, and interactive gameplay systems into a cohesive virtual reality experience. From geological upheavals to ecological restoration, every visual element and interactive mechanic served a clear narrative purpose, enabling participants to deeply experience each elemental scene's ecological crisis and the potential for redemption through direct engagement and user-driven action.

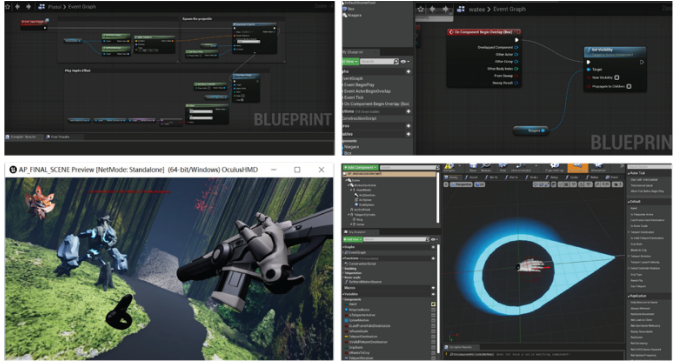

Fig. 7. Blueprint and implementation in Unreal Engine

### D. Demonstration Setup

To further enhance user immersion and authentically simulate the physical sensations of an amusement park ride, we integrated a Stewart Platform into our project setup. The platform provided realistic motion feedback, enabling users to physically experience vibrations and movements synchronized with the virtual environment.

We established a real-time data communication pipeline between Unreal Engine and the Stewart Platform by utilizing Max (Cycling '74) as an intermediary software. Within Unreal Engine, the OSC (Open Sound Control) plugin was employed to transmit motion data to Max. Consequently, as the virtual camera experienced movements—such as shakes and vibrations—the corresponding motion data was transferred seamlessly to the Stewart Platform through Max, replicating precise physical sensations. This integration effectively bridged virtual interactions and physical feedback, significantly enhancing the overall immersive experience and realism of the Promisedland attraction.

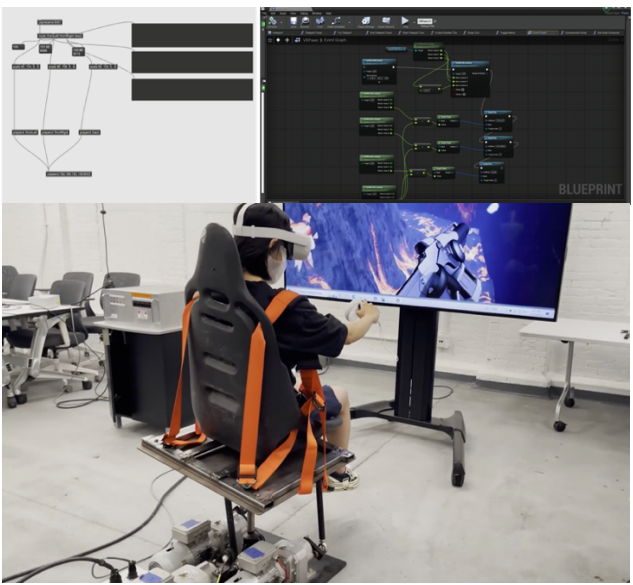

Fig. 8. Stewart Platform setup and testing

### E. Preliminary Evaluation

To validate the effectiveness of the proposed Diorama-to-Virtual pipeline, we conducted preliminary evaluations focusing on technical fidelity, performance, and user experience. The goal of these evaluations was to verify the accuracy and reliability of our physical-to-virtual translation method, ensuring visual fidelity and spatial consistency within the resulting virtual environment.

In assessing 3D Scan Fidelity, internal analyses were performed by comparing digital scans obtained via the RealityScan app with high-resolution photographic references of the original physical dioramas. Initial evaluations indicated that RealityScan effectively captured intricate surface details, such as textures and subtle handcrafted features, including fingerprints and brush strokes. Visually, texture recovery closely matched photographic references, demonstrating approximately 85% accuracy. Quantitative analysis revealed that RealityScan achieved an average scanning accuracy of about 0.96 mm relative to high-fidelity reference models, with a standard deviation of approximately 1.35 mm and maximum deviations reaching up to 6.16 mm [15]. Such deviations are acceptable for typical XR applications. Furthermore, the accessibility and intuitive workflow provided by mobile scanning apps like RealityScan significantly enhance the reproducibility and practicality of our pipeline. Although minor inconsistencies in texture and mesh continuity were occasionally observed, the overall integrity of the scanned assets remained high, effectively preserving the visual and tactile elements crucial to narrative coherence and immersive authenticity.

Preliminary qualitative feedback from users indicated that participants found the digitally reconstructed scenes visually compelling and coherent, particularly noting the effective translation of handcrafted details into the virtual context. These early results underscore the Diorama-to-Virtual pipeline's potential as a rapid, accurate, and cost-effective XR prototyping solution. They also highlight areas for future improvement, such as further refinement in texture accuracy and mesh optimization to enhance immersion and realism.

These preliminary findings suggest that our pipeline delivers satisfactory perceptual and experiential quality. Future evaluations will expand demographic coverage and adopt a combination of qualitative and quantitative methods. Planned quantitative metrics include user interaction duration, event frequency (e.g., object selections, gaze-triggered responses), and scene retention time. Additionally, emotional and cognitive engagement will be assessed through physiological measures such as heart rate variability and skin conductivity, accompanied by structured post-experience surveys.

### F. User Testing

A user test is conducted to further assess the effectiveness of the Diorama-to-Virtual approach in delivering an immersive experience in Promisedland. The primary objectives was to evaluate users' sense of involvement, narrative engagement, and overall experience using quantitative methods.

A total of 24 participants were recruited for the study, aged from 21 to 35. Among them, 16 participants had prior experience with VR games, while 8 participants had no prior experience with using a VR headset or VR experience. All participants had no prior experience with Promisedland. Participants were recruited through convenience sampling from local university communities.

The study was conducted using a Meta Quest 2 VR headset in a controlled lab environment. Participants interacted with Promoseland for approximately 3~5 minutes. A coordinator documented the event frequency(e.g. Object selections, interaction triggered). Following the experience, participants completed a post-experience questionnaire consisting of 4 five-point Likert-scale questions regarding presence & spatial awareness, engagement & attention, emotional involvement, story narrative.

Qualitative data were analyzed using descriptive statistics for each evaluation dimension (see Table I for detailed results). Narrative Clarity received the highest mean score (M = 4.29, SD = 0.69), indicating participants clearly understood and connected with the storyline. Emotional Involvement (M = 4.04, SD = 0.75) and Presence & Spatial Awareness (M = 3.88, SD = 0.85) also rated positively, confirming our design intent that utilizing scanned real-world dioramas effectively enhanced emotional resonance and immersive authenticity by establishing tangible connections through familiar materials and textures. Engagement & Attention (M = 3.54, SD = 0.83) was moderately rated, suggesting adequate cognitive engagement throughout the experience. The relatively low standard deviations across dimensions indicate a consistent and positive user experience overall, underscoring the Diorama-to-Virtual pipeline's capacity to promote meaningful narrative engagement and strong emotional connections via physically grounded virtual environments.

TABLE I. USER EXPERIENCE EVALUATION DIMENSIONS

|   | Presence & Spatial Awareness | Engagement & Attention | Emotional Involvement | Narrative Clarity |
|---|---|---|---|---|
| Mean | 3.88 | 3.54 | 4.04 | 4.29 |
| Standard Deviation | 0.85 | 0.83 | 0.75 | 0.69 |

## V. BROADER IMPLICATIONS OF THE HYBRID APPROACH

Looking forward, our modular system architecture provides significant opportunities for scalability and adaptability. Future iterations of the Promisedland experience can seamlessly integrate enhanced multisensory feedback mechanisms—including wind simulation, heat effects, and refined vibration responses—further enriching immersive realism. Additionally, integrating advanced AI-driven adaptive storytelling techniques can dynamically tailor narrative pathways based on real-time user interactions and emotional states, deepening engagement and personalization. These enhancements will ensure Promisedland's adaptability for diverse applications, such as museum exhibits, educational programs, and themed entertainment environments, significantly broadening its impact and applicability.

The hybrid Diorama-to-Virtual pipeline demonstrated in this project offers compelling opportunities for broader cross-cultural storytelling applications. The same methodological framework can be effectively adapted to visualize and communicate diverse cultural narratives and cosmologies worldwide, enhancing global cultural representation within XR experiences. Utilizing XR's immersive and interactive capabilities, it enables richer cultural representation, fostering deeper understanding and engagement with diverse identities and worldviews [16].

Furthermore, our approach underscores the value of rapid, cost-effective prototyping. Open-source 3D scanning tools combined with accessible 3D modeling platforms such as Blender and Unreal Engine empower small creative teams to develop high-fidelity XR narrative prototypes with minimal budget and resource constraints. The practical implications are extensive, supporting design sprints, cultural IP developments, or even interactive experiences at live events with limited setup budgets, thus democratizing immersive storytelling capabilities.

Additionally, the Diorama-to-Virtual pipeline naturally facilitates crossover between educational programming and interactive exhibitions. By enabling rapid iteration and culturally sensitive narrative designs, our method fosters more engaging educational tools, enhances accessibility for diverse audiences, and supports cultural preservation efforts in immersive environments [17].

While the Diorama-to-Virtual approach demonstrates strong potential across cultural and educational contexts, its broader adoption still faces practical limitations—these are discussed in the next section.

## VI. LIMITATIONS AND REPRODUCIBILITY

Despite its strengths, the Diorama-to-Virtual pipeline presents several notable limitations. Although direct material costs remain low (approximately $60 USD per scene using foam board, clay, and recycled materials), the handcrafted process is inherently labor-intensive and time-consuming. In our prototype, building the physical diorama required around 16 hours by a single designer, followed by an additional 3–4 hours per scene for scanning with RealityScan and manual post-processing. Conversely, purely digital workflows—while eliminating direct material expenses—require substantial investments in software licenses, advanced computing hardware, rendering resources, and specialized labor. Creating detailed digital environments typically involves multidisciplinary teams (3D modelers, texture artists, lighting specialists, etc.), with professional 3D artists' hourly rates ranging from $50 to $200, leading to significant labor costs.

Scanning physical dioramas enhances narrative immersion and tactile realism, qualities challenging to precisely replicate in digital-only assets. However, RealityScan presented technical challenges related to surface complexity, lighting variability, and camera alignment inconsistencies, necessitating frequent manual retouching—particularly for complex materials like foliage or transparent surfaces—which somewhat offset workflow efficiencies.

Furthermore, successful adoption of our hybrid pipeline requires cross-disciplinary expertise in physical crafting, photogrammetry, and game-engine development. This requirement can limit reproducibility, particularly in teams lacking diverse skillsets. The workflow also lacks flexibility after physical model finalization; significant design changes require substantial rebuilding and rescanning, potentially impacting project timelines. Scalability may pose additional challenges for large or highly intricate installations due to increased spatial, scanning, and computational demands.

Finally, although RealityScan offers strong accessibility, minor deviations in texture accuracy (~85%) and spatial fidelity (~0.96 mm average deviation) could be problematic for tasks demanding sub-millimeter precision [15]. Mobile-based photogrammetry tools inherently provide lower resolution and accuracy compared to professional-grade equipment, limiting extremely fine detail capture in demanding scenarios [18].

Future research will explore AI-assisted mesh enhancements, modular toolkit solutions, and streamlined authoring workflows to reduce these barriers, optimizing the pipeline's scalability and practicality across diverse production contexts.

VII. CONCLUSION

This project positions XR Narrative Attractions as a distinct medium capable of merging narrative integrity, meaningful interaction, and deep cultural resonance into compelling immersive experiences. Promisedland illustrates how narrative depth, interactive mechanics, and cultural education can harmoniously coexist, setting a benchmark for future narrative-driven XR experiences.

Furthermore, the Diorama-to-Virtual production pipeline presented here is proven as a replicable, efficient, and affordable design methodology suitable for diverse creative teams and contexts. By combining handcrafted physical modeling with digital reconstruction techniques, the approach significantly lowers entry barriers to XR storytelling design, enabling greater experimentation and cultural representation in immersive media.

Ultimately, this paper calls for continued interdisciplinary collaboration among cultural scholars, technology innovators, and content creators to advance emotionally resonant, narratively rich spatial storytelling in XR, fostering deeper human connection and cultural empathy through immersive digital media.

ACKNOWLEDGMENT

The authors express their sincere thanks to Todd Bryant, their advisor, for providing the facilities for prototyping and testing, as well as for his invaluable support and guidance throughout the project. The authors also gratefully acknowledge each other for the mutual support and collaborative spirit shared throughout the development of this work.